\documentclass{PoS}

\usepackage{cite}
\usepackage{color}

\newcommand{\imag}{{\rm Im}\,}

\newcommand{\eq}[1]{Eq.~(\ref{#1})}
\newcommand{\eqsand}[2]{Eqs.~(\ref{#1}) and (\ref{#2})}
\newcommand{\fig}[1]{Fig.~\ref{#1}}

\newcommand{\lt}{\left}
\newcommand{\rt}{\right}

\title{CP asymmetries in $D$ decays to two pseudoscalars}

\ShortTitle{CP asymmetries in $D\to PP^\prime$ decays}

\author{\speaker{Ulrich Nierste}\\
        Institute for Theoretical Particle Physics (TTP)\\
        Karlsruhe Institute of Technology (KIT)         \\
        76131 Karlsruhe, Germany\\ 
        E-mail: \email{ulrich.nierste@kit.edu}}

\author{Stefan Schacht\\
      Dipartimento di Fisica\\ 
      Universit\`a di Torino \& INFN \\ 
      Sezione di Torino,
      10125 Torino, Italy \\
      E-mail: \email{stefan.schacht@to.infn.it}}

    \abstract{This talk addresses two topics related to CP asymmetries
      in $D$ meson decays to two pseudoscalar mesons.  First I discuss
      how new physics can be distinguished from the Standard Model
      through two sum rules relating three direct CP asymmetries each,
      using the sum rule correlating $D^0\to K^+K^-$, $D^0\to \pi^+
      \pi^-$, and $D^0\to \pi^0 \pi^0$ for illustration. The other sum
      rule involves $a_{CP}^{\mathrm{dir}}(D^+\rightarrow
      \bar{K}^0K^+)$, $a_{CP}^{\mathrm{dir}}(D_s^+\rightarrow
      K^0\pi^+)$, and $a_{CP}^{\mathrm{dir}}(D_s^+\to K^+ \pi^0)$.  The
      second topic is the direct CP asymmetry in the decay $D^0 \to K_S
      K_S$, which is expected to be large in the Standard Model for two
      reasons: Flavor-SU(3) symmetry suppresses the tree amplitude which
      enhances the crucial ``penguin-to-tree'' ratio and the ``penguin''
      amplitude is dominated by the tree-level $W$ boson exchange
      between $c$ and $u$ quarks. We find that $|a_{CP}^{\mathrm{dir}}
      (D^0 \to K_S K_S)|$
      can be as large as $1.1 \%$ in the Standard Model.  We advocate
      $D^0 \to K_S K_S$ as a discovery channel for charm CP violation.}

\FullConference{38th International Conference on High Energy Physics\\
                3-10 August 2016\\
                Chicago, USA\\ and\\
9th International Workshop on the CKM Unitarity Triangle\\
                28  November - 3 December 2016\\
                Tata Institute for Fundamental Research (TIFR), Mumbai,
                India}

\begin{document}

\section{Introduction}
CP asymmetries in the charm system play a special role in the search for
new physics, because they probe flavor-changing transitions among
up-type quarks.  At present, a prime effort of experimental charm
physics is the discovery of CP violation in charm decays. Within the
Standard Model (SM) charm CP asymmetries are small, because the relevant
combination of elements of the Cabibbo-Kobayashi-Maskawa (CKM) matrix is
of order $10^{-3}$.  Precise theoretical predictions
are very difficult and vary by several orders of magnitude 
\cite{Bigi:2011re,Buccella:1994nf, Grossman:2006jg, Artuso:2008vf,
  Petrov:2010gy, 
  Li:2012cfa, Cheng:2012wr, Brod:2011re, Pirtskhalava:2011va,
  Brod:2012ud,Feldmann:2012js,Hiller:2012xm,Franco:2012ck,
  Golden:1989qx, Bobrowski:2010xg}. Since
experimental sensitivities have increased to a level that measurements
probe the range of SM predictions, we need new ideas to draw the
demarcation line between SM and new physics more precisely.
In this talk I discuss two-body weak decays of $D^+,D^0,D_s^+$ mesons
into two pseudoscalar mesons
$P,P^\prime=\pi^0,\pi^{\pm},K_{S,L},K^{\pm}$. My two topics are
\begin{itemize} 
\item[(i)] sum rules relating CP asymmetries in three different $D$
  decays\\ and
\item[(ii)] the CP asymmetry in the decay $D^0\to K_S K_S$. 
\end{itemize}
Topic (i) addresses a test of the Standard Model which will only work,
if at least one of the involved CP asymmetries is measured non-zero.  On
the contrary, topic (ii) is about a discovery mode for charm CP
violation for the case that the Kobayashi-Maskawa phase is the only
source of CP violation. 

Charm decay amplitudes are classified in terms of powers of the 
  Wolfenstein parameter 
\begin{equation}
\lambda\simeq |V_{us}|\simeq|V_{cd}|\simeq
  0.22.
\end{equation}
Amplitudes with  $A\propto
\lt\{
\begin{array}{l}
\lambda^0  \\ 
\lambda^1 \\ 
\lambda^2  
\end{array}\rt\} $ are called 
$\lt\{
\begin{array}{l}
\mbox{Cabibbo-favoured (CF)} \\ 
\mbox{singly Cabibbo-suppressed (SCS)}\\ 
\mbox{doubly Cabibbo-suppressed (DCS)} 
\end{array}\rt\} $. 
In SCS amplitudes $ $ three CKM structures appear,
\begin{displaymath}
\lambda_d=V_{cd}^*V_{ud}, \qquad
\lambda_s=V_{cs}^*V_{us}, \qquad
\lambda_b=V_{cb}^*V_{ub}, 
\end{displaymath}
and CKM unitarity
$\lambda_d+\lambda_s+\lambda_b=0$ is invoked to eliminate 
one combination of CKM elements. A common choice for the decomposition
of an SCS decay amplitude is 
\begin{equation}
\mathcal{A}^{\mathrm{SCS}} \equiv  \lambda_{sd} A_{sd}
 \, - \, \frac{\lambda_b}{2} A_{b} ,
\end{equation}
with
\begin{equation}
\lambda_{sd}=\frac{\lambda_s-\lambda_d}{2} \quad
\mbox{and}\quad
-\frac{\lambda_b}{2}=\frac{\lambda_s+\lambda_d}{2} . 
\end{equation}
In view of
$|\lambda_b|/|\lambda_{sd}|\sim 10^{-3}$ only
$A_{sd}$ is relevant for branching ratios.
Within the SM a non-vanishing direct CP asymmetry involves 
the interference of $\lambda_b A_{b} $ with 
$\lambda_{sd} A_{sd}$. Neglecting quadratic (and higher) terms in 
$\lambda_b/\lambda_{sd}$ the direct CP asymmetry 
reads
\begin{equation}
a_{CP}^{\mathrm{dir}} = \imag \frac{\lambda_b}{\lambda_{sd}} \,
 \imag \frac{A_{b}}{A_{sd}} . \label{eq:acpd}
\end{equation}
Using $\mathcal{A}^{\mathrm{SCS}} \simeq  \lambda_{sd} A_{sd}$ 
and the standard CKM phase convention with (essentially)
$\lambda_{sd}>0$ \eq{eq:acpd} becomes 
\begin{equation}
a_{CP}^{\mathrm{dir}} =  \frac{\imag\lambda_b}{|\mathcal{A}^{\mathrm{SCS}}|}
 \imag \frac{A_{b}}{A_{sd}} |A_{sd}| . \label{eq:acpd2}
\end{equation}
Recalling that $|\mathcal{A}^{\mathrm{SCS}}|$ is determined by the 
well-measured branching ratio of the considered decay  we realize from 
\eq{eq:acpd2} that two non-trivial imputs are needed to   
predict  $a_{CP}^{\mathrm{dir}}$: $|A_b|$ and the relative phase 
between $A_{b}$ and $A_{sd}$. The latter is a CP-conserving (strong)
phase; the CP-violating (weak) phase is $\arg (\lambda_b/\lambda_{sd})$. 

It is not possible to calculate $|A_b|$ and $\arg (A_b/A_{sd})$ from
first principles. The theoretical method of choice in charm physics is
the approximate SU(3)$_F$ symmetry of QCD, which permits to correlate
the amplitudes of different decays with each other. SU(3)$_F$ symmetry
refers to unitary rotations among up, down, and strange fields and would
be exact in the limit $m_u=m_d=m_s$ of equal light-quark masses. The
parameter determining the size of SU(3)$_F$ breaking is
$(m_s-m_d)/\Lambda_{\rm QCD} = {\cal O} (30\%)$, where $\Lambda_{\rm
  QCD}$ is the fundamental scale of QCD.  The actual accuracy of
SU(3)$_F$ symmetry varies among different observables and it is
desirable to include first-order (linear) SU(3)$_F$ breaking.

\section{Sum rules of CP asymmetries}
\begin{figure}[t]
\includegraphics[height=23mm]{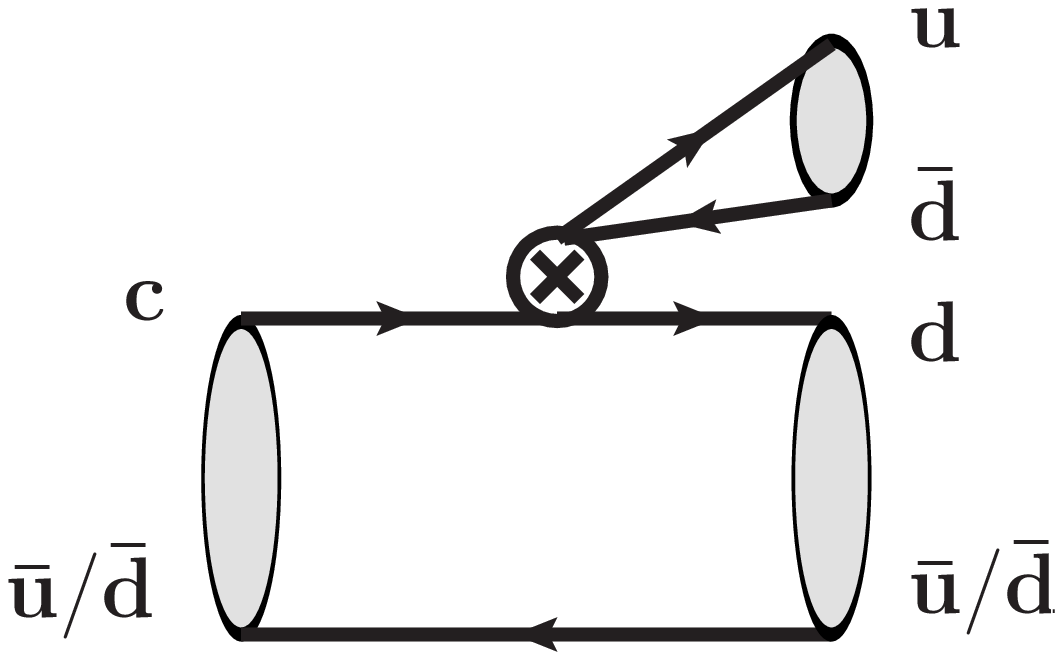}~~
\includegraphics[height=19mm]{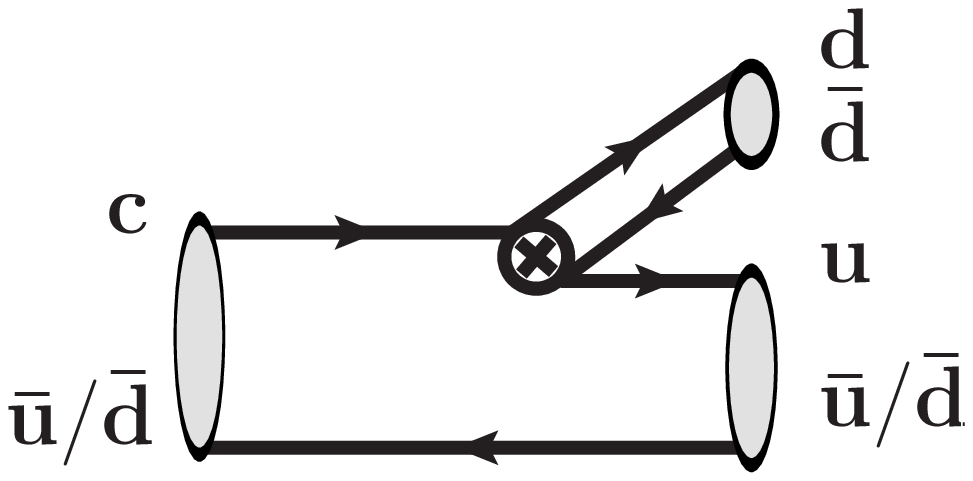}~~
\includegraphics[height=17mm]{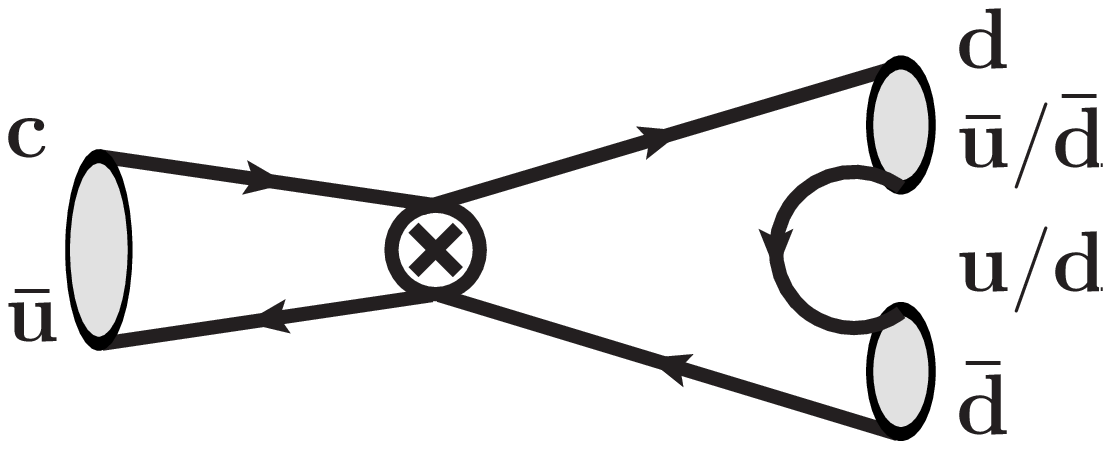}~~
\includegraphics[height=21mm]{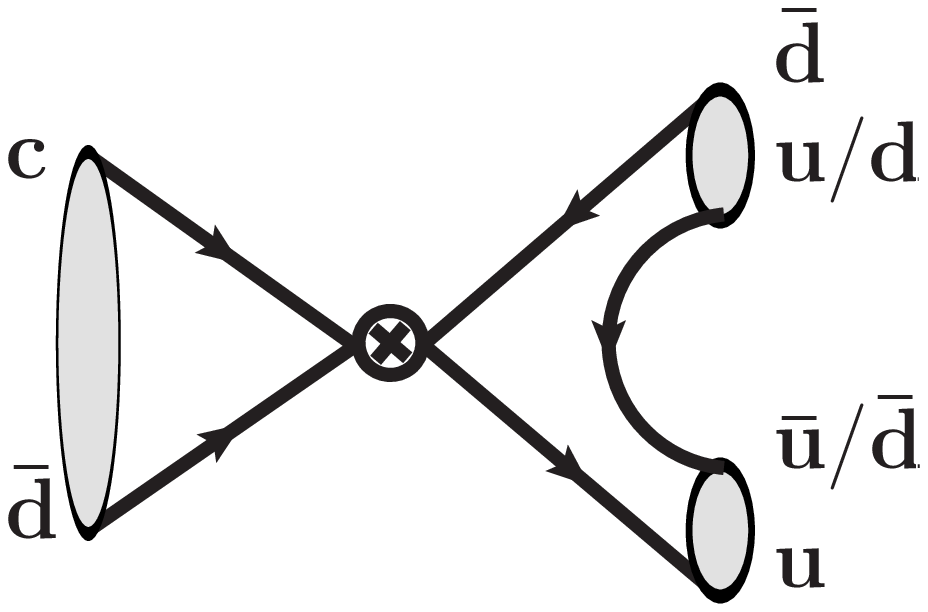}\\[1mm]
\includegraphics[height=20mm]{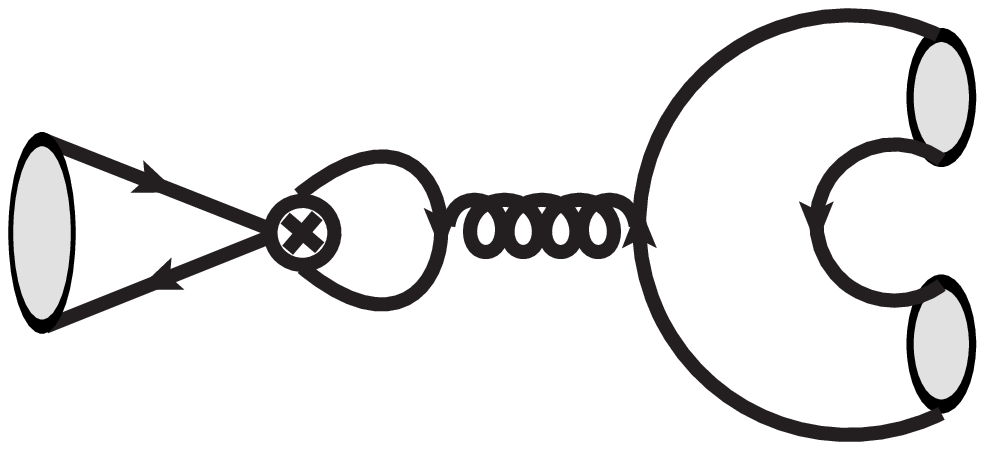}
\hfill
\includegraphics[height=20mm]{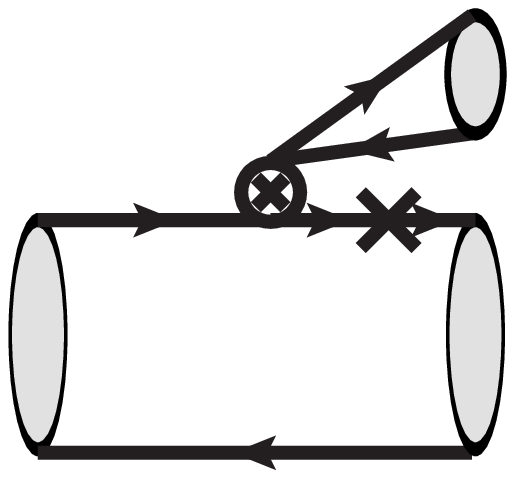}~~
\includegraphics[height=20mm]{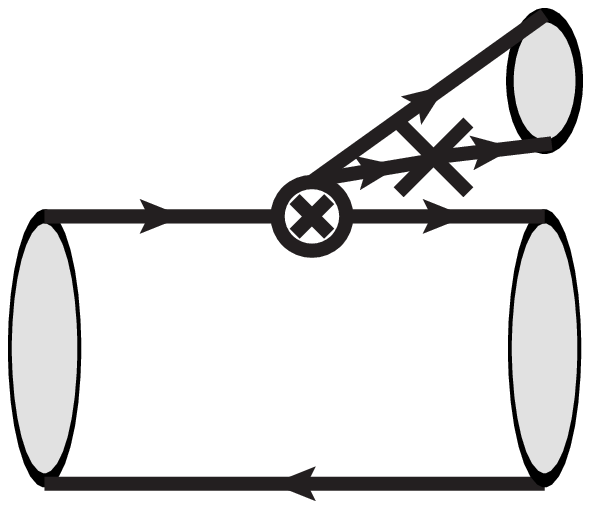}\\
\includegraphics[height=20mm]{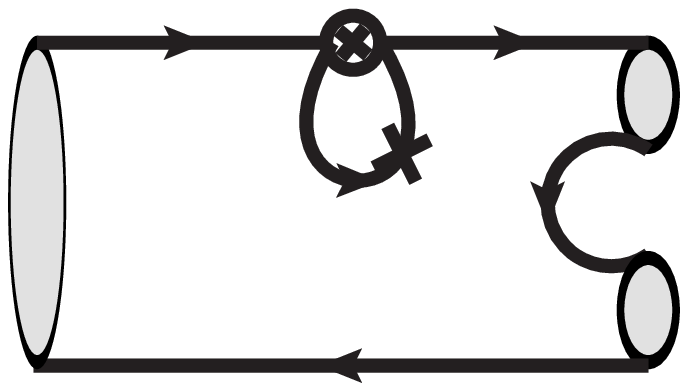}
~~~\parbox[b]{1cm}{\scalebox{1.3}{$\equiv$} \\[1mm]}
\includegraphics[height=20mm]{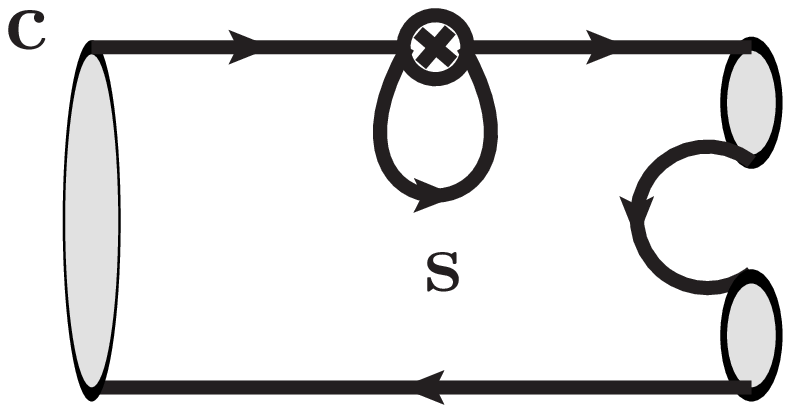}
~~~\parbox[b]{1cm}{\scalebox{1.3}{$-$} \\[1mm]}\hspace{-3mm}
\includegraphics[height=20mm]{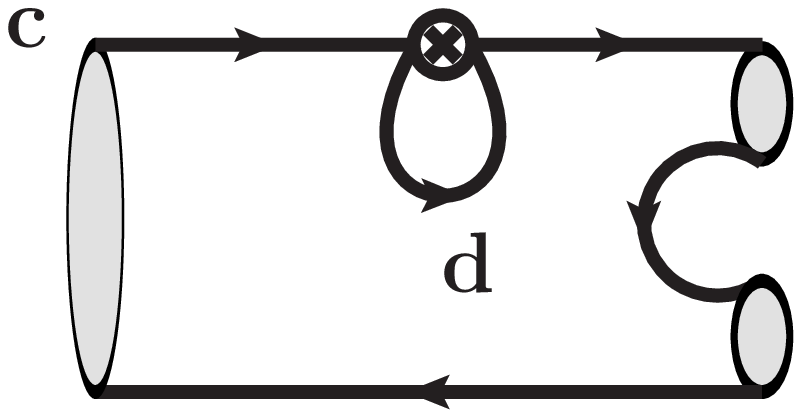}\\
\caption{First row: SU(3)$_F$ limit topological amplitudes ``tree''
  (T), ``color-suppressed tree'' (C), ``exchange'' (E), ``annihilation'' (A).
  Second row: penguin annhilation (PA) diagram and 
  examples of topological amplitudes describing
  SU(3)$_F$-breaking effects (with the cross distinguishing an $s$ from
  a $d$ or $u$ line); the depicted diagrams are $T_1$ and
  $T_2$. Third row: depiction of $P_{\rm
    break}\equiv P_s-P_d$, which is also an SU(3)$_F$-breaking
  amplitude. 
  \label{fig:ta}}
\end{figure}

SU(3)$_F$ analyses can be done in two ways: First, one may express the
physical decay amplitudes in terms of group-theoretical objects, the
reduced amplitudes, which correspond to different representations of
SU(3). It is possible to include first-order SU(3)$_F$ breaking at the
expense of having to deal with more reduced amplitudes (see e.g.\
\cite{Pirtskhalava:2011va,Grossman:2012ry,Hiller:2012xm}). Second, one
can instead express the $D$ decay amplitudes in terms of topological
amplitudes, which are classified by the flavour flow
\cite{Wang:1980ac,Zeppenfeld:1980ex,Chau:1982da}. Also this method
allows for the inclusion of first-order SU(3)$_F$ breaking
\cite{Gronau:1995hm, Muller:2015lua}. The topological amplitudes are
ilustrated in \fig{fig:ta}.
At this stage both methods are mathematically equivalent, {one can
  express the SU(3)$_F$ amplitudes as linear combinations of the
  topological amplitudes and vice versa} 
\cite{Muller:2015lua}.\footnote{{In both  methods one first has to 
remove redundant SU(3)$_F$ or topological amplitudes to obtain a minimal 
set which constitutes a basis for the physical decay amplitudes.}} 
However, the topogical-amplitude method has the
advantage that it can be combined with large-$N_c$ counting
\cite{Buras:1985xv,Buras:1998ra} to sharpen the theoretical predictions.
Here $N_c=3$ is the number of colors and to leading order in the $1/N_c$
expansion the $T$ and $A$ amplitudes can be expressed in terms of form
factors and decay constants.

Armed with this formalism we can study branching fractions and CP
asymmetries of $D\to PP^\prime$ decays. These observables play very
different roles in charm physics: Branching ratios of hadronic charm
decays are ``dull'' tree-level quantities dominated by a single CKM
amplitude and are therefore insensitive to new physics. They are useful
to {test the calculational framework} and experimentally determine
$|A_{sd}|$, which is one of the ingredients to predict {CP asymmetries}.
On the contrary, CP asymmetries of hadronic charm decays are suppressed
by $\imag\frac{\lambda_b}{\lambda_{sd}}=-6\cdot 10^{-4} $ in the {
  Standard Model} and therefore probe {new physics} in flavour
transitions of { up-type} quarks. When exploiting the experimental
information on 16 $D\to PP^\prime$ branching fractions and the $D^0\to
K^\pm \pi^\mp$ strong-phase difference to predict CP asymmetries, one
faces a fundamental problem: CP asymmetries involve topological
amplitudes (equivalently, reduced SU(3)$_F$ amplitudes) which do not
enter branching ratios and are therefore unconstrained by any global fit
to the corresponding data. The most prominent example is the
penguin amplitude: Denoting the penguin with internal quark $q$ by
$P_q$, the branching fractions only constrain the (SU(3)$_F$-breaking)
combination $P_{\rm break}\equiv P_s-P_d$. CP asymmetries, however,
involve not only $P_{\rm break}$ but also $P\equiv P_s+P_d-2 P_b$.

\begin{figure}[t]

\centering 
\includegraphics[width=0.65\textwidth]{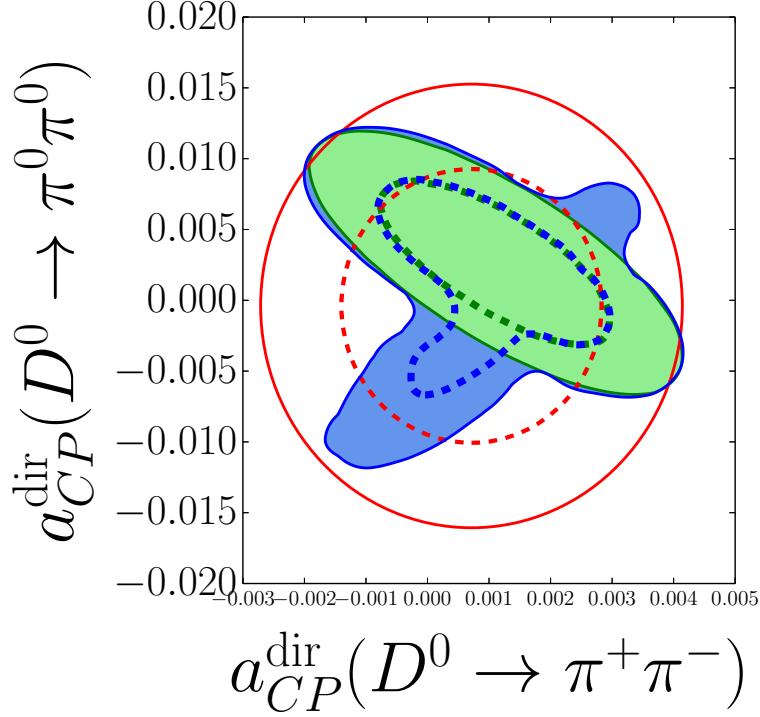}

\caption{The solid (dashed) lines delimit the 95\% (68\%) CL region.
Red: direct measurements. Blue: prediction of sum rule with present
data. Green: sum rule if branching ratios were improved by a factor of
$\sqrt{50}$ compared to today. Plot from Ref.~\cite{Muller:2015rna}.   
\label{fig:plot}}
\end{figure}

Since we cannot predict individual CP asymmetries, we may next try to
predict relations (sum rules) between different CP asymmetries.  
In the limit of exact SU(3)$_F$ symmetry there are two sum rules among
two direct CP asymmetries each \cite{Grossman:2006jg}:
\begin{eqnarray}
  a_{CP}^{\mathrm{dir}}(D^0\rightarrow K^+K^-) +
  a_{CP}^{\mathrm{dir}}(D^0\rightarrow \pi^+\pi^-) &=&
  0\,, \label{eq:su3limit1}\\ 
a_{CP}^{\mathrm{dir}}(D^+\rightarrow
  \bar{K}^0K^+) + a_{CP}^{\mathrm{dir}}(D_s^+\rightarrow K^0\pi^+) &=&
  0\,. \label{eq:su3limit2}
\end{eqnarray}
In Ref.~\cite{Grossman:2013lya} it has been shown that there are no sum
rules among CP asymmetries which hold to first order in SU(3)$_F$ breaking.
Can we improve on \eqsand{eq:su3limit1}{eq:su3limit2} anyway?
To this end consider, for example, 
\begin{equation}
   A_{sd} (D^0\to \pi^+\pi^-)
           =-T-E+P_{\mathrm{break}}, \quad\qquad
  A_b (D^0\to \pi^+\pi^-)=T+E+P+PA , \label{eq:ex1}
\end{equation}
which entails $A_b (D^0\to \pi^+\pi^-)=-A_{sd} (D^0\to \pi^+\pi^-)+
P_{\mathrm{break}}+P+PA$. ($PA$ is defined analogously to $P$.) 
Then \eq{eq:acpd} reads 
\begin{equation}
 a_{CP}^{\mathrm{dir}} (D^0\to \pi^+\pi^-) = 
   \imag \frac{\lambda_b}{\lambda_{sd}} \,
   \imag \frac{P_{\mathrm{break}}+P+PA}{A_{sd}(D^0\to \pi^+\pi^-)} . \label{eq:acpd3}
\end{equation}
In the SU(3)$_F$ limit the corresponding expression for
$a_{CP}^{\mathrm{dir}} (D^0\to K^+ K^-) $ is indeed equal in magnitude
and opposite in sign. One can next use the global branching-ratio
analysis of Ref.~\cite{Muller:2015lua} to determine all ingredients of $
a_{CP}^{\mathrm{dir}} (D^0\to \pi^+\pi^-)$ and $a_{CP}^{\mathrm{dir}}
(D^0\to K^+ K^-) $ including first-order SU(3)$_F$ breaking, except for
$P+PA$ which is unconstrained. I.e.\ we are left with two quantities
depending on two unknowns, which are real and imaginary part of
$P+PA$. In order to make a prediction we therefore need a third a
quantity, $ a_{CP}^{\mathrm{dir}} (D^0\to \pi^0\pi^0)$. By eliminating
$P+PA$ we find the desired sum rule.  The global fit \`a la
Ref.~\cite{Muller:2015lua} determines all topological amplitudes
entering $A_{sd}$ for the three decay modes (which are $T$, $E$, $C$,
$T_{1,2}$, $E_{1,2}$, and $P_{\rm break}$), so that the troublesome
SU(3)$_F$-breaking terms causing $B(D^0\to K^+K^-)\neq B(D^0\to
\pi^+\pi^-)$ are taken care of. However, $P+PA$ cannot be treated beyond
the SU(3)$_F$ limit.  \fig{fig:plot} shows the impact of the sum
rule. Similarly, we can improve \eq{eq:su3limit2} to a sum rule
involving also the third decay mode $D_s^+\to K^+ \pi^0$. With current
data the sum rule has much larger errors than the one for $D^0\to
K^+K^-,\pi^+\pi^-,\pi^0\pi^0$.

{In the discussion after the talk the issue of final state
  interactions was raised. In our formalism all amplitudes $T$,
  $E,\ldots$ are understood to comprise all effects of the strong
  interaction, including final-state rescattering. There are several 
 attempts in the literature to separate final-state interactions from other 
 QCD effects, see
 e.g. \cite{Buccella:1990sp,Chua:2005dt,Biswas:2015aaa}. A commonly used
 ansatz for some decay amplitude $A_j\equiv A(D\to f_j)$ 
is \cite{Chua:2005dt,Biswas:2015aaa} 
 \begin{equation}
 A_j = \sum_{k} S^{1/2}_{jk} A_k^{(0)}, \label{eq:rsc}  
\end{equation}
where $S^{1/2}$ is the matrix describing the scattering of the final state
$f_k$ into $f_j$ and $A_k^{(0)}$ denotes the amplitude
in the absence of any rescattering. Here we remark that in general the
establishment of a formula like \eq{eq:rsc} requires assumptions on the
underlying dynamics, otherwise the definition of $ A_k^{(0)}$ and
$S^{1/2}_{jk}$ is ambiguous.  Remarkably, one can prove \eq{eq:rsc} for
$K\to \pi\pi$ decays using solely the isospin symmetry of the strong
interaction (\emph{Watson's theorem} \cite{Watson:1952ji}). In the case
of $D$ or $B$ decays one can justify \eq{eq:rsc} for the
\emph{absorptive} part\footnote{{In the limit of real CKM elements the
  absorptive and dispersive parts of a weak decay amplitude simply equal
  the imaginary and real parts of the amplitude, respectively.}} of $A_j$
with the optical theorem. However, the \emph{dispersive} part does not
obey \eq{eq:rsc} with the $S^{1/2}_{jk}$  inferred from the optical
theorem, because it also involves a summation over virtual (off-shell)
states and not just the real states label by $k$ in \eq{eq:rsc}.  
We emphasize that a decomposition like \eq{eq:rsc} is not necessary 
for an analysis employing only SU(3)$_F$ symmetry and $1/N_c$ counting,
since it permits to treat all strong-interaction effects (whether
stemming from final-state interaction or not) on the same
footing.}

\section{CP asymmetry in $D^0\to K_S K_S$}
$D^0\to K_S K_S$ has the special feature that $A_{sd}$ vanishes in the
SU(3)$_F$ limit. The smallness of $|A_{sd}|$ enhances $A_b/A_{sd}$, so
that $a_{CP}^{\mathrm{dir}}$ in \eq{eq:acpd} is expected to be larger
than in other decays. Concerning statistical errors this observation
does not help, because the gain in statistical significance is cancelled
by the smaller number of events, which scale with the branching ratio
proportional to $|A_{sd}|^2$. In our case we have $B(D^0\to K_S
K_S)=(1.7\pm 0.4)\cdot 10^{-4}$. Still, a larger CP asymmetry may help
to fight systematic errors. Another special feature of the considered
decay mode is way more interesting: The numerator $A_b$ in \eq{eq:acpd}
receives contributions from the exchange diagram $E$, so that the CP
asymmetry persists even if the loop-induced amplitudes $P$ and $PA$ 
(which induce the CP asymmetries in essentially all other decay modes)
turn out to be tiny. Moreover, the global fit of Ref.~\cite{Muller:2015lua}
points to a large $E$ and definitely excludes $E=0$. The sensitivity 
to $E$ stems from the feature that in $D^0\to K_S K_S$ the transitions
$c \bar u \to s\bar s$ and $c \bar u \to d\bar d$ can interfere, because 
both $s\bar s$ and  $d\bar d$ can hadronize into a $K_S K_S $ pair.  
In Ref.~\cite{Nierste:2015zra} we find 
\begin{eqnarray}
  -1.1 \cdot 10^{-2} \leq a_{CP}^{\mathrm{dir}} \leq 1.1 \cdot 10^{-2}.
\end{eqnarray}
This number assumes that the CP asymmetry related to Kaon mixing is
properly subtracted.  Unfortunately, the global fit to all $D\to
PP^\prime$ data presently does not rule out that
$|a_{CP}^{\mathrm{dir}}|$ is much smaller than $1.1 \cdot 10^{-2}$. One
source of uncertainty is the strong phase $\arg (A_b/A_{sd})$ which is
currently unconstrained.  To eliminate this source of uncertainty, one
must also measure the mixing-induced CP asymmetry \cite{Nierste:2015zra},
with a time-dependent measurement or through CP-tagged
decays. Currently experiments determine
\begin{equation}
A_{CP} = a_{CP}^{\mathrm{dir}} - A_{\Gamma} \frac{\langle
  t\rangle}{\tau},
\end{equation}
where $\langle t\rangle$ is the average decay time and $\tau$ is the
$D^0$ lifetime. $A_{\Gamma}$ involves the mixing-induced CP asymmetry
and is small, because $D^0$ mesons oscillate very slowly.  The
experimental results are
\cite{Bonvicini:2000qm,Aaij:2015fua,Abdesselam:2016gqq}
\begin{eqnarray}
A_{CP}^{\mathrm{CLEO~2001}} &=& -0.23 \pm 0.19, \qquad
A_{CP}^{\mathrm{LHCb~2015}} = -0.029 \pm 0.052 \pm 0.022, \nonumber\\
\qquad  A_{CP}^{\mathrm{Belle~2016}} &=& -0.0002 \pm 0.0153 \pm 0.0017.
\nonumber
\end{eqnarray} 

\section{Summary}
CP asymmetries in $D$ decays involve topological amplitudes which are
not constrained by fits to branching ratio data.  These can be
eliminated by forming judicious combinations (sum rules) of several CP
asymmetries.  Within the limits of expected SU(3)$_F$ breaking in
penguin (and penguin annihilation) amplitudes these sum rules probe 
new physics. Within the Standard Model the direct CP asymmetry in
    $D^0 \to K_S K_S$ can be as large as 1.1\%.
    $a_{CP}^{\mathrm{dir}} (D^0\to K_S K_S)$ is dominated by 
    the exchange diagram, which involves no loop suppression.
    We advocate $D^0 \to K_S K_S$ as a  potential 
    discovery channel for charm CP violation.

\section*{Acknowledgements}
I thank the organisers for inviting me to this talk. 
The presented work is supported by BMBF under contract  05H15VKKB1.  Parts
of the computations were performed on the {NEMO cluster of
  Baden-W\"urttemberg (framework program bwHPC).}

\end{document}